\documentclass[12pt]{article}
\usepackage{amsmath,amssymb,epsf,latexsym,enumerate,cite,shadow,array,color}
\usepackage[ps,dvips,matrix,arrow,frame,import,curve,color]{xy}
\UseCrayolaColors

\newif\iffigs\figstrue

\DeclareFontFamily{U}{rsf}{}
\DeclareFontShape{U}{rsf}{m}{n}{
  <5> <6> rsfs5 <7> <8> <9> rsfs7 <10-> rsfs10}{}
\DeclareMathAlphabet\Scr{U}{rsf}{m}{n}

\def\pplogo{\vbox{\kern-\headheight\kern -29pt
\halign{##&##\hfil\cr&{
\ppnumber}\cr\rule{0pt}{2.5ex}&\ppdate\cr}
}}
\makeatletter
\def\ps@firstpage{\ps@empty \def\@oddhead{\hss\pplogo}%
  \let\@evenhead\@oddhead 
}
\def\maketitle{\par
 \begingroup
 \def\thefootnote{\fnsymbol{footnote}}
 \def\@makefnmark{\hbox{$^{\@thefnmark}$\hss}}
 \if@twocolumn
 \twocolumn[\@maketitle]
 \else \newpage
 \global\@topnum\z@ \@maketitle \fi\thispagestyle{firstpage}\@thanks
 \endgroup
 \setcounter{footnote}{0}
 \let\maketitle\relax
 \let\@maketitle\relax
 \gdef\@thanks{}\gdef\@author{}\gdef\@title{}\let\thanks\relax}
\makeatother

\def\IZ{\mathbb{Z}}
\def\IR{\mathbb{R}}

\newcommand{\mc}{\mathcal}
\def\mcA {{\mc A}}
\def\mcC {{\mc C}}
\def\mcG {{\mc G}}
\def\mcL {{\mc L}}

\newcommand{\eq}[1]{Eq.~(\ref{eq:#1})}

\def\one{{\hbox{ 1\kern-.8mm l}}}

\def\tr{{\rm tr\,}}

\begin{document}
\setcounter{page}0
\def\ppnumber{\vbox{\baselineskip14pt
\hbox{RUNHETC-2008-09}}}
\def\ppdate{May 2008} \date{}

\title{\LARGE Kinetic terms in warped compactifications\\[10mm]}

\author{Michael R. Douglas$^{1,2,\&}$
 and Gonzalo Torroba$^1$\\
\\
$^1$NHETC and Department of Physics and Astronomy\\
Rutgers University\\
Piscataway, NJ 08855--0849, USA\\[5mm]
$^2$Simons Center for Geometry and Physics, Stony Brook NY 11790, USA \\[5mm]
$^\&$I.H.E.S., Le Bois-Marie, Bures-sur-Yvette, 91440 France\\
\small{{\tt mrd@, torrobag@physics.rutgers.edu}}}

{\hfuzz=10cm\maketitle}

\def\Large{\large}
\def\LARGE{\large\bf}

\vskip 1cm

\begin{abstract}
We develop formalism for computing the kinetic terms of 4d fields in
string compactifications, particularly with warping. With the help of
the Hamiltonian approach, we identify a gauge dependent inner product
on the compactification manifold which depends on the warp factor. It
is shown that kinetic terms are associated to the minimum value of the
inner product over each gauge orbit. We work out the kinetic term for
the complex modulus of a deformed conifold with flux, {\it i.e.} the
Klebanov-Strassler solution embedded in a compact Calabi-Yau manifold.
Earlier results of a power-like divergence are confirmed qualitatively
(the kinetic term does contain the main effect of warping) but not
quantitatively (the correct results differ by an order one
coefficient).
\end{abstract}

\vfil\break

\tableofcontents

\vskip 2mm
\section{Introduction}\label{sec:intro}    

One of the central problems of string/M theory is to find consistent
compactifications and work out their four dimensional low energy
descriptions.  Most work starts with the 10d or 11d supergravity
theory and does Kaluza-Klein reduction, and then considers stringy and
quantum effects as corrections depending on small parameters.  We can
refer to a regime in which such an expansion is good as a
``supergravity limit.''  Using duality, many strong coupling limits
can be reformulated as other weakly coupled supergravity limits.  But
on general grounds one expects other ``order one coupling'' regimes to
exist, and there has been much effort to understand them, by summing
instantons, using holomorphy, interpolating between different weakly
coupled regimes, etc.

While this is an important goal, the supergravity limits already
realize a great deal of interesting physics, and could be better
understood.  Indeed, our experience has been that a key to the general
problem has been to identify mathematical structures present in the
supergravity limit, which persist in the general case.  This was the
case for mirror symmetry, both closed string (variation of Hodge
structure) and open string (categorical and $A_\infty$ structure,
stability conditions).  Thus, our goals include both developing
practical calculational techniques, and to find such structures.

In the present work, we focus on the problem of computing kinetic
terms.  Our immediate motivation was the study of type IIb flux
compactification carried out in \cite{dst,stud} along lines initiated
in \cite{gkp}.  These are warped compactifications, and Kaluza-Klein
reduction in such backgrounds is subtle, with various incorrect and
incomplete results in the literature.  One reason for this is that, 
in following the standard approach of substituting the Kaluza-Klein
ansatz into the Lagrangian, one finds that one needs ``compensator''
fields~\cite{gray, gm}, which are difficult to solve for explicitly, and do not
(at least to us) suggest any clear physical or mathematical intuition
for the results.

As it turns out, a fairly direct route to the kinetic terms is to
derive them in a Hamiltonian framework.  The reason is that the system
has constraints associated to gauge redundancies, while the physical
degrees of freedom become manifest in the Hamiltonian formulation.
While this does not completely eliminate the need to discuss
compensators, it does provide a much clearer picture of why they arise
and how to deal with them.

Perhaps the simplest way to explain the main point is to realize that
the kinetic terms for metric moduli originate from a metric on the
space of metrics, but the usual expression for this metric is gauge
dependent.  A mathematically natural \cite{Singer:1981xw} and
physically correct \cite{manton} way to fix this ambiguity is to
require that the metric fluctuations be orthogonal to gauge
transformations.  However, when one says ``orthogonal,'' one has
implicitly used the ten-dimensional metric, in a way which sees the
warp factor.  This is the point at which warping changes the usual
discussion.

\subsection{General Problem}

We consider a $D$-dimensional theory of gravity coupled to matter,
{\it e.g.} a supergravity. A vacuum solution is a solution of the
equations of motion which at long distances ``looks like'' a
$d$-dimensional space $M$ with maximal symmetry, {\it i.e.} Minkowski
space, AdS or dS.  In general it will be a product or warped product
of $M$ with an $n=D-d$-dimensional compactification space (or internal
space) $X$, possibly with other nonzero fields consistent with maximal
symmetry ({\it i.e.} scalars, components of vector fields in $X$,
etc.).

We use $x^\mu$ and $y^i$ to denote coordinates on $M$ and $X$ respectively.
For definiteness we will sometimes take $D=10$ and $d=4$, but our
considerations will not depend on this.

Suppose there is a family of vacuum solutions of the $D$-dimensional
equations of motion, with parameters $u^I$.  Thus we can write
$g_{MN}(y;u)$, $\phi(y;u)$, and so forth. To analyze the dynamics of
these moduli $u^I$, we might try to find a family of ``approximate
solutions'' of the equations of motion, obtained by taking the
parameters to slowly vary on $M$~\cite{manton}:
\begin{equation} \label{eq:naive-ansatz}
g_{MN}(y,u(x)) .
\end{equation}

The kinetic terms are then the terms in the $d$-dimensional effective
Lagrangian of the form
\begin{equation} \label{eq:kinetic}
\int d^dx\ \sqrt{g} g^{\mu\nu}\,
 G_{IJ}(u)\, \partial_\mu u^I \partial_\nu u^J ,
\end{equation}
obtained by substituting \eq{naive-ansatz} into the $D$-dimensional
action, integrating over $X$ and identifying these terms.\footnote{
The correct action may require a boundary term to cancel boundary
terms in the variation, for example the Gibbons-Hawking-York term
in general relativity.}
Note that to compute \eq{kinetic}, we
need to allow ``off-shell'' $u(x)$ ({\it i.e.} $\partial^2 u\ne 0$).

However, this direct approach can become complicated.  The first sign
of this is that in general, the ansatz \eq{naive-ansatz} does not
solve the ten-dimensional equations of motion, even when $u(x)$ solves
the four-dimensional massless field equations.  One may need a more
general ansatz depending on derivatives $\partial u, \partial^2 u$,
etc.  Further subtleties arise from gauge invariance. We will see how
this happens and its consequences in examples.

\subsection{Summary}

We start in section \ref{sec:ym} with the example of Yang-Mills
theory, which is used to illustrate in a simple setup many of the
subsequent points. Then in section \ref{sec:gr} we construct a
Riemannian metric on the space of metrics, with the help of the
Hamiltonian of General Relativity. This metric is used in section
\ref{sec:kinetic} to construct kinetic terms arising from 10d (warped)
backgrounds preserving 4d maximal symmetry. We prove that metric
fluctuations should be orthogonal to gauge transformations associated
to the full warped metric. This turns out to be equivalent to
minimizing the value of their inner product over each gauge orbit.

In section \ref{sec:string}, the previous formalism is applied to
string compactifications. We first discuss the case of a Calabi-Yau
manifold, where the metric for complex and K\"ahler moduli is
recovered. The harmonic gauge choice generally considered in the
literature is identified as a dynamical constraint. Next the more
interesting case of conformal Calabi-Yau compactifications is
analyzed; these correspond to type IIb supergravities with BPS branes
and fluxes. Compensating fields are identified with Lagrange
multipliers of the Hamiltonian. Their role is to set metric
fluctuations into harmonic gauge with respect to the full warped
metric. We find a fairly simple expression for the field space metric
in terms of warped metric fluctuations. Upon rewriting this in terms
of the underlying Calabi-Yau moduli we verify the expression recently found
in~\cite{stud}.

Finally, in section \ref{sec:ks} we compute the metric for the complex
modulus $S$ of the warped deformed conifold. We find a power-like
divergence $|S|^{-4/3}$ that agrees with the analysis done
in~\cite{dst}. Both results differ, however, by a numerical
coefficient. The reason for this is that before it was not known how
to construct fluctuations orthogonal to gauge transformations.

\section{Yang-Mills theory}\label{sec:ym}

We start 
with the simple case of a $U(1)$ field $A_M$ with field strength $F_{MN}$.
We suppose that there are a family of solutions of
$$
D^i F_{ij} = 0
$$
on $X$, parameterized by coordinates $u^I$.
For example, if $X$ is a torus, every flat connection
is a solution, and the $u^I$ might be the
holonomy associated to a basis of $H^1(X,\IZ)$.

We take as the ten-dimensional action
\begin{equation} \label{eq:tendim-action}
S = \ldots
- \frac{1}{4}\int d^4x \sqrt{g_4} \int d^6y \sqrt{g_6}
  g^{MN} g^{PQ} F_{MP} F_{NQ} .
\end{equation}
Naively we then set $A_\mu=0$ and write
$$
F_{\mu i} = \partial_\mu A_i(y;u(x)) - \partial_i A_\mu(y;u(x))
 = \frac{\partial A_i}{\partial u^I} \partial_\mu u^I ,
$$
and substitute this into the action, to obtain \eq{kinetic} with
\begin{equation} \label{eq:maxwell-metric}
G_{IJ} = \int d^6y \sqrt{g_6} g^{ij} 
 \frac{\partial A_i}{\partial u^I}
 \frac{\partial A_j}{\partial u^J} .
\end{equation}
However, on reflection, there must be a subtlety in this procedure.
In defining our moduli space of solutions $A_i(y;u)$, nowhere did
we specify a gauge for $A_i$.  Two solutions which are related by
gauge transformations on $X$,
$$
\delta A_i = \partial_i \epsilon ,
$$
are equally good from the point of view of $X$.
On the other hand, the expression \eq{maxwell-metric} is not gauge
invariant, so the kinetic terms will depend on which of the gauge
equivalent solutions we take. Since \eq{tendim-action} was gauge invariant 
in ten dimensions,
we must have made an error.  

The error was the assumption that $A_\mu=0$
for all of these solutions.
Let us look at the ten dimensional equations of motion.  These
can be written as
\begin{equation} \label{eq:ten-d-eom}
0 = D^\mu F_{\mu\nu} + D^i F_{i\nu} ; \qquad
0 = D^\mu F_{\mu j} + D^i F_{i j}  .
\end{equation}
We substitute the ansatz $A_i(y;u(x))$ and require that there is
no four-dimensional gauge field, $F_{\mu\nu}=0$.  This sets
$$
A_\mu(x,y)=\Omega(y) \partial_\mu f(x)
$$
where $\Omega(y)$ and $f(x)$ are still undetermined functions.
 
To find $A_\mu$, we use the first equation of motion, which becomes
$0 = \partial^i F_{i\nu}$, {\it i.e.}
\begin{equation} \label{eq:comp-equation}
\partial^i \partial_\nu A_i = \partial^i \partial_i A_\nu .
\end{equation}
In general, the left hand side is nonzero, so we will have $A_\nu\ne 0$.
However a simple way to make the left hand side zero is to require
\begin{equation} \label{eq:gauge-condition}
0 = \partial^i \frac{\partial A_i}{\partial u^I} ,
\end{equation}
{\it i.e.} the fluctuations are taken in harmonic gauge.
More generally, solving \eq{comp-equation} produces an $A_\nu$ which 
is the parameter of the ``compensating gauge transformation'',
\begin{equation} \label{eq:compensating-gauge}
A_\mu(x,y)=\Omega_I(y) \partial_\mu u^I(x)\;,\;
 \partial^i \partial_i \Omega_I=\partial^i \;\frac{\partial A_i}{\partial u^I}\,.
\end{equation}
Defining
\begin{equation} \label{eq:physical0m}
\delta_I A_i:=\frac{\partial A_i}{\partial u^I}-\partial_i \Omega_I\,,
\end{equation}
we see that the effect of $\Omega$ is to put $\delta_I A_i$ back into 
harmonic gauge.

In general, it is hard to explicitly solve \eq{comp-equation} for
the compensator field $A_\nu$.  However, to compute the kinetic term,
we do not need to do this, rather we just need to impose the condition
\eq{gauge-condition}.  

\subsection{Metric for Yang-Mills connections}

One can straightforwardly generalize the above to nonabelian gauge
fields.  There is also a simple geometric interpretation of the final
result, which leads immediately to the metric both for Yang-Mills and
for gravitational configurations. 

Note that \eq{gauge-condition} is the condition that the variation
$\delta_I A$ is orthogonal in the metric \eq{maxwell-metric} to all
the gauge directions.
This is a natural mathematical condition and leads to
a unique definition of the metric \cite{Singer:1981xw}.

Let $\mcA$ be the set of possible (smooth) gauge potentials on
$\IR^3$, and $\mcG$ be the group of all gauge transformations over
$\IR^3$.  The four-dimensional physical configuration space is then
the quotient (or orbit space) $\mcC \equiv \mcA/ \mcG$.

Given a metric $g_{ij}$ on $\IR^3$, there is a natural metric
on $T\mcA$,
\begin{equation} \label{eq:ymc-metric}
({\dot A},{\dot A}) = \int d^3x\ \sqrt{g} g^{ij} \;\;
\tr \left({\dot A_i(x)} {\dot A_j(x)}\right) .
\end{equation}

Given a path $c(t)$ in $\mcC$, we would like to define a natural
Riemannian metric $H$ on $\mcC$, which
can be used in a particle action as~\cite{manton}
\begin{equation}\label{eq:ym-metric}
S[c]=\int dt\,\frac{1}{2}\,H(\dot c, \dot c)\,.
\end{equation}
Since actually one works with paths $A_i(t)\,\in \,\mcA$, the basic
requirement is that $S[c]$ should be independent of the way $c(t)$ is
lifted to $\mc A$. This can be accomplished
by projecting the tangent vector
$\dot A_i(t)$ on the subspace orthogonal to gauge transformations
in the metric \eq{ymc-metric}.
Thus, let $\Pi_i$ be this projection,
\begin{equation}\label{eq:ym-projector}
\Pi_i(\dot A):= \dot A_i-D_i (1/D^2) D_j \dot A_j\;,\;\Pi_i(D_k \lambda)=0\,.
\end{equation}
The natural metric on $\mcC$ is then 
\begin{equation} \label{eq:ym-metric2}
H(\dot c, \dot c)=\int d^3x\,{\rm tr}\big( \Pi_i(\dot A)\, \Pi_i(\dot A)\big)\,.
\end{equation}

From a physics point of view, $\Pi_i(\dot A)$ is the electric field
$F_{0i}$ after eliminating $A_0$ by using the Gauss law. Equivalently, the
projector is given by the nonabelian version of the zero mode
\eq{physical0m} after solving for the compensator $\Omega$.  Substituting
into the $E^2$ terms of the Yang-Mills action, one obtains \eq{ym-metric}.

There are several other formulations of the same result.  One is to
regard the configuration space $\mcA$ as a $\mcG$-bundle over the
space of gauge orbits.  The projection \eq{ym-projector} then defines
a preferred notion of ``parallel transport'' on this bundle, making
\eq{ymc-metric} unambiguous.  The metric \eq{ym-metric} is then gauge
invariant, in the sense that it is derived from a gauge invariant
notion of parallel transport.

Another formulation is to note that, since the metric \eq{ymc-metric}
is positive definite, evaluating it with the gauge directions
projected out is the same as evaluating it on the gauge representative
which \emph{minimizes} its value.

\subsection{Relation to Hamiltonian formulation}

A slightly different way of reducing to gauge invariant variables is
to go to the Hamiltonian formulation.  We recall that, since the time
derivatives $\partial_0 A_0$ do not appear in the action, the 
$A_0$ component of the vector potential plays the role of a Lagrange
multiplier, which is conjugate to the Gauss law,
$$
S = \ldots + \int A_0 D_i E^i .
$$
One can then enforce the Gauss law as a constraint on the initial
data $(A_i,E^i)$, which is preserved under Hamiltonian evolution.

This is a particular example of ``symplectic reduction'' with respect
to a symmetry group $G$.  Starting with a phase space $M$ with a
symplectic structure $\omega(u,v)$, one identifies ``moment maps''
$\mu$ which are ``Hamiltonians'' generating the infinitesimal action
of $G$.  One can then show that the reduced phase space
$$
\{x\in M:\mu(x)=0\} / G
$$
carries a symplectic structure.  

In the Yang-Mills example, $G=\mcG$, and $M$ is the direct product of 
the space $\mcA$ of connections $A_i(x)$
with the space of electric field strengths $E^i(x)$.
It carries the symplectic structure
$$
\omega(A,E) = \int d^3x\ \tr\left( A_i(x) E^i(x) \right) .
$$ The moment maps for $\mcG$ are then $\mu=D_i E^i$.  Thus, the Gauss
law constraint is the natural partner of the gauge condition in this
construction as well. Since the $E^2$ terms in the Hamiltonian are
gauge invariant, they are single valued on the reduced phase space,
resulting in the same metric \eq{ym-metric}.

Physically, we can use this formulation by considering a configuration
in which the moduli $u^I$ are linearly varying with time.  The metric
is then the energy density of this configuration, and the Hamiltonian
framework provides a direct way to compute this.  Since the phase
space does not contain time-like components of vector potentials,
there is no possibility for a ``compensator field'' $A_0$ to enter;
rather the mixed equations of motion such as \eq{ten-d-eom} 
are solved implicitly in this framework.

In general, the result of this prescription will depend on the
initial choice of symplectic structure on field space.  However in
field theory there is usually a unique local candidate for this structure.

\section{General relativity}\label{sec:gr}

In this section we consider the problem of constructing a natural
Riemannian metric on the space of metrics. This will be done by using
the Hamiltonian formulation of general relativity, which is
well-suited for extracting the kinetic terms in a general case. At the
end of the section we present a simple example where the kinetic terms
are obtained via the usual Lagrangian approach, so that both
perspectives may be compared.

\subsection{Metric on the space of metrics}\label{sec:hamilton}

The problem may be formulated as follows. Consider a $D$-dimensional
manifold equiped with a metric $g_{MN}(x)$, $M, N=0,\ldots, D$. In
many cases of interest the metric satisfies certain background
equations of motion.  For example, in pure Einstein gravity it is
Ricci flat.  However these equations depend on the theory, and thus we
will not make use of them in this section.

We identify a time coordinate $t=x^0$; then $\Sigma$ denotes the
space-like surface $t=0$ and $h_{MN}$ is the pull-back of $g_{MN}$ to
$\Sigma$. Let $\mcA$ be the set of all such possible Riemannian
metrics $h_{MN}$, and $\mcG$ the corresponding diffeomorphisms. Our
aim is to identify a Riemannian metric $H$ on $\mcA/\mcG$ and then
for each path $c(t) \,\in \, \mcA/\mcG$ introduce a natural action
\begin{equation}\label{eq:c-action}
S[c]=\int dt\,\frac{1}{2} H(\dot c, \dot c)\,.
\end{equation}
Following the previous discussion it will now be shown how this arises
from the Hamiltonian formulation for GR~\cite{adm,wald}.

\vskip 1.5mm

One starts by prescribing initial value conditions on a $D-1$
dimensional space-like surface $\Sigma_0$, with metric $h_{MN}$.
Denoting its unit normal vector by $n_N$,
\begin{equation}
h_{MN}=g_{MN}+n_M n_N\,.
\end{equation}
The equations of motion produce the time evolution $\Sigma_0 \to \Sigma_t$, and
the physical degrees of freedom are $h_{MN}$ and not $g_{MN}$. The remaining
components, denoted by $\eta_N$, are determined in terms of the ``dual''
vector $t^M$, which satisfies
$$
(g_{tt})^{1/2}=-g_{MN}t^M n^N\,.
$$
Recall the gauge choice $t=x^0$; also, $g_{tt}=-g_{00} > 0$. Then,
$$
\eta^N=h^{NM} \eta_M=t^N-(g_{tt})^{1/2} n^N\,.
$$ 
The geometrical interpretation is that the time evolution $\Sigma_0
\to \Sigma_t$ given by the vector field $t^N$ can be decomposed into a
normal direction $n^N$ plus a tangential shift $\eta^N$. The dynamics
is encoded in the extrinsic curvature,
\begin{equation}\label{eq:extrinsicK}
K_{MN}:=\frac{1}{2} \mathcal L_n h_{MN}=\frac{1}{2}
  (g^{tt})^{1/2}(\dot h_{MN}-D_N \eta_M -D_M \eta_N)\,,
\end{equation}
where $D_N$ is the covariant derivative on $\Sigma$, compatible with
$h_{MN}$. The lagrangian density takes the form
\begin{equation}\label{eq:LG}
\mathcal L_G= \sqrt{-g_D} \Big( R^{(D-1)}+K_{MN} K^{MN}-K^2\Big)\,.
\end{equation}

In terms of these variables, the canonical momentum reads
\begin{equation}\label{eq:pi}
\pi^{MN}=\frac{\partial \mathcal L_{G}}{\partial \dot h_{MN}}=
h^{1/2} (K^{MN}-h^{MN}K)\,,
\end{equation}
from which we obtain the Hamiltonian density,
\begin{equation}\label{eq:HG}
\mathcal H_G= \sqrt{-g_D} \Big( -R^{(D-1)} +h^{-1} \pi^{MN} \pi_{MN}-
\frac{1}{D-2} h^{-1} \pi^2\Big)-2 h^{1/2} \eta_N D_M (h^{-1/2} \pi^{MN})
\end{equation}
The shift vectors $\eta^N$ are Lagrange multipliers which
enforce the constraints
\begin{equation}\label{eq:pi-constr}
D_N (h^{-1/2}\pi^{NM})=0\,.
\end{equation}
After satisfying this we can set $\eta^N=0$, as usual in constrained
Hamiltonian systems.

\vskip 1.5mm

The Riemannian metric on $\mc A/\mc G$ corresponds to the kinetic term
of the Hamiltonian density. Given a path $c_{MN}(t) \in \mcA/\mcG$ we
introduce a lift $h_{MN}(t)$ to $\mcA$; to the tangent vector $\dot
h_{MN}$ we associate the ``projection'' $\pi^{MN}(\dot h)$ defined in
\eq{pi}. The metric on the space of metrics becomes
\begin{equation}\label{eq:Hmetric}
H(\dot c,\,\dot c)=\int d^{D-1}x\,\sqrt{-g_D}\Big( h^{-1} \pi^{MN} \pi_{MN}-
\frac{1}{D-2} h^{-1} \pi^2 \Big)
\end{equation}
and the action is given by \eq{c-action}. The constraint \eq{pi-constr} implies that $\pi^{MN}(\dot h)$ is orthogonal to gauge transformations,
$$
H(\mcL_v \pi,\,\pi)=0\,.
$$
Actually, $\pi^{MN}$ itself is a projector $\mcA \to \mcA/\mcG$:
$$
\pi^{MN}(\mcL_v \dot h)=0\,.
$$ 
The proof is analogous to the YM case \eq{ym-projector}, and is
based on eliminating the Lagrange multipliers $\eta_N$. We conclude
that the Hamiltonian approach to GR yields a natural Riemannian metric
\eq{Hmetric} on $\mcA/\mcG$.

\subsection{Unwarped solutions}\label{subsec:unwarped}

In simple cases it is still possible to compute kinetic terms
using the Lagrangian formulation, as we now discuss in an example.
Consider a family of six dimensional Ricci-flat manifolds $X$ with
metric $g_{ij}(y; u)$. Examples are Calabi-Yau manifolds, with $u^I$
parametrizing complex and K\"ahler moduli. The ten dimensional
background is taken to be the unwarped product $M \times X$ with
metric
\begin{equation}\label{eq:background}
ds^2=g_{\mu \nu}(x) dx^\mu dx^\nu+ g_{ij}(y; u) dy^i dy^j\,.
\end{equation}

Promoting the moduli to fields $u^I(x)$ fibers $X$ over $M$, but only
through the implicit dependence of the moduli on the space-time
coordinates. As in the Maxwell case, just replacing $u \to u(x)$ into
\eq{background} doesn't give a consistent $D$-dimensional solution.
To satisfy $G_{MN}=0$, we consider the following ansatz including a
compensating field $B_i$:
\begin{equation}\label{eq:metric-ansatz}
ds^2=g_{\mu \nu}(x) dx^\mu dx^\nu
+2 B_{Ij}(y)\partial_\mu u^I dy^j dx^\mu+ g_{ij}(y; u(x)) dy^i dy^j\,.
\end{equation}
It has been pointed out~\cite{gm} that an extra compensator term of
the form $K_I(y) \partial_\mu \partial_\nu u^I\,dx^\mu dx^\nu$ may
also be needed. However, we will show that $B_{Ij}$ is only defined
modulo a total derivative term, which can be used to set $K_I=0$.

The components of the Einstein tensor, up to two space-time derivatives, read
\begin{equation}\label{eq:Gmunu}
G_{\mu \nu}=(\partial_\mu \partial_\nu u^I-g_{\mu \nu}\Box u^I)
\big[-\frac{1}{2}\frac{\partial g}{\partial u^I}
  +\nabla^jB_{Ij}\big]
\end{equation}
\begin{equation}\label{eq:Gmui}
G_{\mu i}=\frac{1}{2} \partial_\mu u^I \nabla^j
\big(\nabla_i B_{Ij}-\nabla_j B_{Ii}+\frac{\partial g_{ij}}{\partial u^I}
-g_{ij} \frac{\partial g}{\partial u^I}\big)
\end{equation}
\begin{equation}\label{eq:Gij}
G_{ij}=-\frac{1}{2}\Box u^I
\big[\frac{\partial g_{ij}}{\partial u^I}-\nabla_i B_j
-\nabla_j B_i \big]\,\,,
\end{equation}
where the trace part is
$$
\frac{\partial g}{\partial u^I}:=g^{ij} \frac{\partial g_{ij}}{\partial u^I}\,.
$$ 

A consistent ten dimensional solution requires $G_{\mu i}=0$, which
fixes $B_{Ij}$, up to a total derivative $\partial_j K_I$. Then we
have to require that $G_{\mu \nu}=0$, off-shell for $u(x)$, which
determines the previous function $K_I$:
\begin{equation}\label{eq:K}
\nabla^j B_{Ij}=\frac{1}{2}\frac{\partial g}{\partial u^I}\,.
\end{equation}
Using \eq{K} to eliminate $\partial_I g$, \eq{Gmui} can be rewritten 
more suggestively as
\begin{equation}\label{eq:B}
\nabla^i \Big[\frac{\partial g_{ij}}{\partial u^I} -\nabla_i B_j
-\nabla_j B_i\Big]=0\,.
\end{equation}

Plugging these results in the Einstein-Hilbert action, the action
up to two space-time derivatives is of the form \eq{kinetic}, with
field space metric
\begin{equation}\label{eq:unwarpedG}
G_{IJ}(u)=\frac{1}{4}\int d^6y \sqrt{g_6} \,g^{ij} g^{kl}\, \delta_I g_{ik}\,
\delta_J g_{jl}
\end{equation}
where
\begin{equation}\label{eq:udeltag}
\delta_I g_{ij}:=\frac{\partial g_{ij}}{\partial u^I}-\nabla_i B_j-\nabla_i B_j\,.
\end{equation}
The role of the ten dimensional constraints is to set $\delta_I
g_{ij}$ in the transverse traceless gauge,
\begin{equation}\label{eq:harmonic-gauge}
\nabla^i \,\delta_I g_{ij}=0\;,\;g^{ij} \delta_I g_{ij}=0\,.
\end{equation}

This example shows how the metric compensators repackage into a
``physical'' zero mode $\delta_I g_{ij}$ which is orthogonal to
diffeomorphism transformations. Their effect can be simply summarized
in the requirement that the zero mode has to be in the transverse
traceless gauge. The upshot from this example is that harmonic gauge
is not a choice, but rather a dynamical constraint.

\section{Kinetic terms in general compactifications}\label{sec:kinetic}

The most general $D$-dimensional metric consistent with
$d$-dimensional maximal symmetry is
\begin{equation}\label{eq:warped-product}
ds^2=e^{2A(y;\,u)}\,\hat g_{\mu \nu}(x) dx^\mu dx^\nu+
g_{ij}(y;u) dy^i dy^j\,.
\end{equation}
This is a warped product of a maximally symmetric space $M$ with metric
$\hat g_{\mu \nu}$ and an arbitrary compactification manifold $X$ with
metric $g_{ij}$. The internal manifold depends on parameters $u^I$ and
the aim is to find their kinetic terms. This applies to all 
supergravity compactifications preserving 4d maximal symmetry.

We will assume here that $g_{ij}$ does not have exact isometries, as is
the case in CY manifolds. This simplifies the analysis, since there
are no gauge fields coming from the off-diagonal fluctuations $\delta
g_{\mu m}$. There is a mass gap and $\delta g_{\mu m}$ are associated
to massive spin 1 fields, which we choose not to excite. In a more
complete treatment, one should describe how such fields combine with
the graviton modes (and scalars from the internal manifold) to yield
massive spin 2 degrees of freedom.

The situation is a particular case of that discussed in the previous
section, where the path $c(t)$ corresponds to promoting $u^I$ to spacetime
fields. Since the 4d part $\hat g_{\mu \nu}$ is fixed, the metric on the space
of metrics should now reduce to a metric on the parameter space $\{u^I\}$.
We will not assume that $g_{ij}$ is Ricci-flat; rather, it satisfies certain
background equations of motion (for instance, including fluxes). The advantage
of the Hamiltonian approach is that the identification of the kinetic
term does not require analyzing such equations.

Once the $u^I$ are allowed to fluctuate, we have to include
compensators $B_{Ij}$,
\begin{equation}\label{eq:warped-comp}
ds^2=e^{2A(y;\,u)}\,\big(\hat g_{\mu \nu}(x) dx^\mu dx^\nu+2B_{Ij}(y) \partial_\mu u^I \,dx^\mu dy^j\big)+
g_{ij}(y;u) dy^i dy^j\,.
\end{equation}
In the Lagrangian approach, the compensators are fixed by solving the
equations of motion at linear order in velocities. Once this is done,
the kinetic terms may be extracted from the equations which are
quadratic in space-time derivatives.

Here the system will be analyzed from a Hamiltonian point of view; for
simplicity, we take $\partial_\mu u^I=\delta^0_\mu \dot u^I$.\footnote{Recall that the
difference between $g_{MN}$ and $h_{MN}$ is that the latter only
includes space-like components.} The kinetic term for the moduli $u^I(t)$ is obtained by plugging the corresponding time-dependent metric $\dot h_{MN}= \dot u^I (\partial h_{MN}/\partial u^I)$ in \eq{Hmetric}. In the linearized approximation the extrinsic curvature $K_{MN}$
and canonical momentum $\pi_{MN}$ are both proportional to $\dot u^I$, so we can write
\begin{equation}\label{eq:physh}
K_{MN}= \frac{1}{2}\,(g^{tt})^{1/2}\, \dot u^I\,\delta_I h_{MN}\;,\;h^{-1/2} \pi_{MN}=\frac{1}{2}\,(g^{tt})^{1/2}\, \dot u^I\,\delta_I \pi_{MN}
\end{equation}
and the factors of $(g^{tt})^{1/2}/2$ have been extracted for later convenience. The coefficients $\delta_I h_{MN}$ and $\delta_I \pi_{MN}$ are given by
\begin{equation}
\delta_I h^{MN}=\frac{\partial h^{MN}}{\partial u^I}-D^M \eta^N_I
-D^N \eta^M_I
\end{equation}
\begin{equation}\label{eq:pi-0m}
\delta_I \pi^{MN}=\delta_I h^{MN}-h^{MN}\,h^{PQ}\,\delta_I h_{PQ}
\end{equation}
where we have expanded $\eta^N=\dot u^I\,\eta^N_I$.

The relation between the Lagrangian and Hamiltonian approach is that
the compensators coincide with the Lagrange multipliers $\eta_{M}$,
\begin{equation}\label{eq:shift}
\eta_{I\mu}=0\;,\;\eta_{Ij}=e^{2A}\,B_{Ij}(y)\,.
\end{equation}
The advantage of the Hamiltonian formulation is that they appear explicitly as nonpropagating fields, whose only role is to impose the constraints
\begin{equation}\label{eq:Dconstr}
D^N\big((g^{tt})^{1/2}\delta_I \pi_{MN} \big)=0\,,
\end{equation}
which imply that the physical variations are orthogonal to gauge transformations. We remind the reader that $D_N$ is the covariant derivative compatible with the space-like metric $h_{MN}$. The kinetic term derived from the Hamiltonian \eq{Hmetric} reads
\begin{eqnarray}\label{eq:inner2}
H&=&\frac{1}{4}\,\dot u^I\,\dot u^J\, \left(\int d^{D-1}x\, \sqrt{-g_D}\,g^{tt}\,\big[\delta_I \pi_{MN} \,\delta_J \pi^{MN}-\frac{1}{D-2} \,\delta_I \pi\,\delta_J \pi \big] \right)\nonumber\\
&=& \frac{1}{4}\,\dot u^I\,\dot u^J\, \left(\int d^{D-1}x\, \sqrt{-g_D}\,g^{tt}\,\delta_I \pi_{MN} \,\delta_J h^{MN}\right) \,.
\end{eqnarray}
This is the gravitational analog of the kinetic term $p\, \dot q$ in particle mechanics.
 
Let us now prove that \eq{Dconstr} is equivalent to \emph{minimizing}
the inner product over each gauge orbit. Under a gauge transformation
$$
\delta_I h^{MN} \to \delta_I h^{MN}-D^N v_I^M-D^M v_I^N\,,
$$
the change in the inner product \eq{inner2} is
\begin{equation}
-2\int d^{D-1} \sqrt{-g_D}\,g^{tt}\, v_I^M\, D^N \left[(g^{tt})^{1/2}\left(\delta_J \pi_{MN}+\mc L_v\,\delta_J \pi_{MN}\right)\right]\,.
\end{equation}
Demanding that the gauge parameter minimizes this expression, we find
\begin{equation}
D^N \left[(g^{tt})^{1/2}\left(\delta_J \pi_{MN}+\mc L_v\,\delta_J \pi_{MN}\right)\right]=0\,,
\end{equation}
thus reproducing the prescription given in \eq{Dconstr}.

\subsection{Four dimensional expression}

To compactify over the internal manifold one would in principle need
to know the warp factor and then extract the variation $\partial_I
A$. These are complicated functions determined by the background
equations of motion. But interestingly, the constraints \eq{pi-constr}
fix $\delta_I A$ in terms of $g^{ij}\,\delta_I g_{ij}$: from
$$
0=D^\mu(\delta_I \pi_{\mu\nu})=-\partial_\nu\big(2\,e^{-2A}\,\delta_I e^{2A}+g^{ij}\,\delta_I g_{ij} \big)\,,
$$
we obtain
\begin{equation}\label{eq:deltaA}
\delta_I e^{2A}=-\frac{1}{2}\,e^{2A}\,g^{ij} \,\delta_I g_{ij}\,.
\end{equation}
This implies that $\delta \pi_{\mu\nu}=0$, while the warp factor variation may be eliminated from $\delta \pi_{ij}$ yielding
\begin{equation}\label{eq:delta-bg}
\delta_I \pi_{ij}=\delta_I g_{ij}+\frac{1}{d-2}
g_{ij}\,g^{kl}\delta_I g_{kl}\,.
\end{equation}
The internal part of the constraint sets
\begin{equation}\label{eq:constr-0m}
D^N(e^{-A}\delta_I \pi_{Nj})=0\,,
\end{equation}
where $e^{-A}$ comes from $(g_{tt})^{-1/2}$, and it is important to remember that the connection is defined with respect to the full warped metric. To rewrite this in terms of 6d variables, notice that
$$
D^\mu(e^{-A} \delta_I \pi_{\mu j})= 3 \,e^{-A}\,\partial^k A\;\delta_I \pi_{kj}
$$
where we used the fact that $\pi_{\mu \nu}=0$ and $h^{\mu \nu} \Gamma^k_{\mu\nu}=-3 \partial^k A$. Then (\ref{eq:constr-0m}) becomes
\begin{equation}\label{eq:constr6d}
g^{ij}\nabla_i(e^{2A}\delta_I \pi_{jk})=0\,.
\end{equation}

With these results, the general formula for the kinetic terms
is\footnote{We are ignoring the overall factor $M_{P,D}^{D-2}$; also
the correct normalization of the $d$-dimensional Ricci term would
introduce a factor of $1/Vol(X)$ in the field space metric.}
\begin{equation}
S_{kin}=\frac{1}{2} \int d^d x\,\sqrt{-\hat g_d}\, \hat g^{tt}\, \dot u^I \dot u^J \,G_{IJ}(u)
\end{equation}
with
\begin{equation}\label{eq:hamilton-G}
G_{IJ}(u)=\frac{1}{4} \int d^{D-d}y \sqrt{g_{D-d}}\,e^{2A}\, \delta_I g_{ij}\,\delta_J
\pi^{ij}\,.
\end{equation}
The warp factor dependence comes from 
$\sqrt{-g_d}\,g^{tt}=\sqrt{-\hat g_d}\,\hat g^{tt}\,e^{2A}$. From this expression it becomes clear that \eq{constr6d} is simply the condition
that the physical variation $\delta_I \pi_{ij}$ is orthogonal to gauge transformations. The effects of the compensators are
summarized in this prescription.

\subsection{Effect of compensators}\label{subsec:comp-effects}

The Hamiltonian approach shows that the effect of the compensators is
to make the metric fluctuations orthogonal to gauge
transformations. In general it is simpler to compute the ``naive''
zero modes just by taking derivatives $\frac{\partial g_{ij}}{\partial
u^I}$. The metric associated to these fluctuations is
\begin{equation}
G^0_{IJ}=\frac{1}{4} \int d^{D-d}y \sqrt{g_{D-d}}\,e^{2A}\, \Big(
\frac{\partial g_{ij}}{\partial u^I}\,\frac{\partial g^{ij}}{\partial u^J} -
\frac{1}{D-2} \,\frac{\partial g}{\partial u^I}\,\frac{\partial g}{\partial u^J}\Big)\,,
\end{equation}
which is a gauge-dependent quantity because in general $\partial_I
g_{ij}$ is not orthogonal to gauge transformations.

Starting from $G^0_{IJ}$ we can ask what is the effect of the ``compensating gauge transformation''
\begin{equation}\label{eq:deltag}
\delta_I g_{ij}=\frac{\partial g_{ij}}{\partial u^I}-\nabla_i \eta_{Ij}-\nabla_j \eta_{Ii}
\end{equation}
which projects down to $\mathcal A/\mathcal G$. More concretely, we
are interested in analyzing $G_{IJ}-G_{IJ}^0$, which may be shown to
be
\begin{equation}
G_{IJ}-G_{IJ}^0=\frac{1}{4}\,\int d^{D-d}y\,\sqrt{g_{D-d}}\,e^{2A}\,\eta_{Ij}\,\nabla_i\left(\frac{\partial \pi^{ij}}{\partial u^J}\right)+(I \leftrightarrow J)\,.
\end{equation}

Let's first derive the explicit projector analogous to the expression
\eq{ym-projector} for nonabelian Yang-Mills theories. From
\eq{constr-0m}, the compensating fields satisfy the equation
\begin{equation}\label{eq:comp}
\big( g_{ij}\,\nabla^k \nabla_k+2 \nabla_i \nabla_j+R_{ij}\big)\,\eta_I^j=\nabla^k\left(\frac{\partial g_{ki}}{\partial u^I}\right)
\end{equation}
plus the relation \eq{deltaA} which fixes possible residual gauge
transformations preserving \eq{comp}. Defining the operator
$$
\mc O_{ij}:=g_{ij}\,\nabla^k \nabla_k+2 \nabla_i \nabla_j+R_{ij}\,,
$$
formally the compensators are given by
\begin{equation}
\eta_I^i=(\mc O^{-1})^{ij}\,\nabla^k(\frac{\partial g_{kj}}{\partial u^I})\,.
\end{equation}
In this way,
\begin{equation}
\delta_I g_{ij}=\frac{\partial g_{ij}}{\partial u^I}-\nabla_i\,(\mc O^{-1})_{jl}\,\nabla_k\left(\frac{\partial g^{kl}}{\partial u^I} \right)+(i \leftrightarrow j)\,.
\end{equation}

We conclude that the effect of the compensators on the metric is
\begin{equation}\label{eq:compG}
G_{IJ}-G_{IJ}^0=\frac{1}{2}\,\int d^{D-d}y\,\sqrt{g_{D-d}}\,e^{2A}\,
\nabla_i\left(\frac{\partial g^{ij}}{\partial u^I}\right)\,\mc O^{-1}_{jl}\,\nabla_k\left(\frac{\partial g^{kl}}{\partial u^I}\right)
\,.
\end{equation}
This is the term responsible for minimizing the metric over each
gauge orbit. A different compensator choice would imply that the gauge
directions are not projected out, giving a larger result.

\section{Application to string compactifications}\label{sec:string}

The Hamiltonian derivation of the field space metric \eq{hamilton-G}
holds quite generally. In particular supersymmetry is not assumed and
the details of the matter sector (fluxes, branes, etc.) are not
needed.

Of course, given supersymmetry, one can exploit its constraints.  For
instance, for $\mc N=2$ supersymmetries the metric for chiral
superfields may be obtained from that of the vector superpartners in
the $\mc N=2$ multiplet, which enter quadratically in the 10d
action. Already for $\mc N=1$ susy, deriving the moduli kinetic terms
by dimensionally reducing the 10d action supersymmetry is a very
involved task, as was shown in~\cite{stud}. The main obstacle is the
correct implementation of the constraints, which arise from the $(0M)$
components of Einstein equations.

On the other hand, we have shown how the kinetic terms arise more
naturally from the GR Hamiltonian. In this section, some simple
examples of type II compactifications will be analyzed from this point
of view.

\subsection{Calabi-Yau manifolds}\label{subsec:cy}

To gain intuition we begin by discussing Calabi-Yau compactifications,
both from the Hamiltonian and Lagrangian viewpoint. An unwarped
Calabi-Yau compactification corresponds to
\begin{equation}\label{eq:CYm}
ds^2= g_{\mu \nu}(x) dx^\mu dx^\nu+g_{ij}(y) dy^i dy^j\,,
\end{equation}
where $g_{ij}$ is a Ricci flat K\"ahler metric. Holomorphic
coordinates are denoted by $z^a$, $a=1,2,3$, so that the K\"ahler form
is $J=ig_{a \bar b}\,dz^a \wedge d\bar z^b$. The metric moduli space
splits into complex structure deformations $S^\alpha \,\delta_\alpha
g_{ab}$, and K\"ahler deformations $\rho^r\, \delta_r g_{a \bar b}$.

The Hamiltonian analysis may be applied straightforwardly to this
case. The space-time components of the constraint \eq{Dconstr} imply
that the metric fluctuations must be traceless, while the internal
components tell us that the fluctuations are in harmonic gauge:
\begin{equation}\label{eq:CYconstr}
g^{ij}\,\delta_I g_{ij}=0\;,\;\nabla^i(\delta_I g_{ij})=0\,,
\end{equation}
with $I$ running over $(\alpha,\,r)$. These conditions were a choice
in the 6d approach of Candelas and de la Ossa~\cite{candelas}, but
here they emerge as constraints of the 10d Hamiltonian picture.  This
occurs as follows. Starting from a zero mode $\partial g_{ij}/ \partial u^I$ in
some arbitrary gauge, the compensators are equivalent to a
diffeomorphism transformation $\partial_I g_{ij} \to \delta_I
g_{ij}=\partial_I g_{ij}-\nabla_{(i}\,B_{Ij)}$ which point to point
imposes the transverse-traceless constraints. The metric
\eq{hamilton-G} gives, after reintroducing the Planck mass,
\begin{eqnarray}\label{eq:GCY}
G_{\alpha \bar \beta}&=&\frac{1}{4V_{CY}}\,\int d^6y \sqrt{g_6}\,g^{a \bar c} g^{b \bar d}\,\delta_\alpha g_{ab}\,\delta_\beta g_{\bar c \bar d}\nonumber\\
G_{rs}&=&\frac{1}{4V_{CY}}\,\int d^6y \sqrt{g_6}\,g^{a \bar c} g^{b \bar d}\,\delta_r g_{a\bar d}\,\delta_s g_{b \bar c}\,.
\end{eqnarray}

Let us explain briefly how the zero modes are actually computed,
because this will be necessary to understand conformal Calabi-Yau
compactifications. Since \eq{CYm} is a solution without sources,
starting from a given background value $g^0_{ij}$, the zero modes are
solutions to
\begin{equation}
R_{ij}(g^0+\delta g)=0\,.
\end{equation}
Recalling the linearized expression for the Ricci tensor~\cite{wald}
$$
\delta R_{ij}=-\frac{1}{2}\nabla^k \nabla_k \delta g_{ij}-\frac{1}{2}\nabla_i \nabla_j \delta g+ \nabla^k \nabla_{(i} \delta g_{j)_k}\,,
$$
the zero mode fluctuations satisfy
\begin{equation}\label{eq:deltaR}
-\frac{1}{2}\nabla^k \nabla_k \delta g_{ij}-\frac{1}{2}\nabla_i \nabla_j \delta g+R_{k(ij)l} \delta g^{kl}+\frac{1}{2}\big(\nabla_i \nabla^k \delta g_{kj}+\nabla_j \nabla^k \delta g_{ki} \big)=0\,.
\end{equation}

Next, imposing the gauge $\nabla^i \delta g_{ij}=0$, 
the trace part can be set to zero and one is left with
\begin{equation}\label{eq:lich}
-\frac{1}{2}\nabla^k \nabla_k \delta g_{ij}+R_{k(ij)l} \delta g^{kl}=0\,.
\end{equation}
This gauge-fixed version of $\delta R_{ij}=0$ is the Lichnerowicz
laplacian on Ricci-flat manifolds.\footnote{If the Ricci-tensor
doesn't vanish there is an extra term proportional to $R_{ik} \delta
g_j^{\phantom{1}k}$. However, the Einstein equation would also include
a source piece.} On a K\"ahler manifold the only nonzero components of
the Riemann tensor are $R_{a \bar b c \bar d}$ up to permutations,
which implies that the zero modes of mixed ($\delta g_{a \bar b}$) and
pure ($\delta g_{ab}$) type separately verify this equation.

\subsection{Conformal Calabi-Yau case}

At the next level of complexity, we consider an internal manifold
which is a conformal Calabi-Yau, with the conformal factor given by
the inverse of the warp factor,
\begin{equation}\label{eq:conformal}
ds^2=e^{2A(y)} \eta_{\mu \nu}(x)\,dx^\mu dx^\nu+e^{-2A(y)}\,\tilde g_{ij}(y)\, dy^i dy^j\,,
\end{equation}
where $\tilde g_{ij}$ is the CY metric. These type IIb backgrounds preserve $\mc N=1$ susy, and the warp factor is generated by BPS sources~\cite{gkp}.

In terms of the unwarped fluctuations $\delta_I \tilde g_{ij}$, the constraint \eq{deltaA} sets
\begin{equation}\label{eq:constr1}
\delta_I A=\frac{1}{8}\,\tilde g^{ij}\delta_I \tilde g_{ij}\, ;
\end{equation}
this fixes the 4d gauge redundancies. Now $\delta \pi_{ij}$ given
in \eq{delta-bg}, becomes the warped harmonic combination
\begin{equation}\label{eq:bg2}
\delta_I \pi_{ij}=e^{-2A} (\delta_I \tilde g_{ij}-\frac{1}{2} \tilde g_{ij}\, \delta_I \tilde g)\,.
\end{equation}
The constraint coming from $D_M \pi^{Mj}=0$ sets
\begin{equation}\label{eq:constr2}
g^{ik}\nabla_i(e^{2A}\delta_I \pi_{kj})=\tilde g^{ik}\tilde \nabla_i \big(\delta_I \tilde g_{kj}-\frac{1}{2} \tilde g_{kj}\,\delta_I \tilde g \big)-4\tilde g^{ik}\,\partial_i A\,\delta_I \tilde g_{kj}=0\,.
\end{equation}
Finally, replacing \eq{bg2} into the Hamiltonian expression \eq{hamilton-G}, we arrive to the warped moduli space metric
\begin{equation}\label{eq:warped-G}
G_{IJ}(u)=\frac{1}{4V_W} \int d^6y \sqrt{\tilde g_6}\,e^{-4A}\, \tilde g^{ik} \tilde g^{jl}\,
\delta_I \tilde g_{ij}\,\delta_J \tilde g_{kl}\,.
\end{equation}

These results agree with those in~\cite{stud}, which were obtained by
dimensionally reducing the action. In that approach, the compensators
were gauged away; in the Hamiltonian formalism they arise as Lagrange
multipliers which can always be set to zero. Furthermore, the rather
complicated constraint in the r.h.s. of \eq{constr2} has a simple
interpretation in terms of the full metric with conformal and warp
factors, $\nabla^i (e^{2A}\delta_I \pi_{ij})=0$. The present derivation
suggests that the natural metric fluctuations are $\delta \pi_{ij}$
instead of $\delta A$ and $\delta \tilde g_{ij}$ separately.

The presence of a nontrivial warp factor has important effects on the
moduli dynamics. \eq{constr1} implies that the fluctuations acquire a
nonzero trace part proportional to $\delta_I A$; on the other hand,
\eq{constr2} imposes a gauge which is different from the harmonic
condition. Therefore, although the fields $u^I$ are the same as in the
unwarped case (so that we still have complex and K\"ahler moduli), the
internal wavefunctions that support them have changed. From
\eq{deltag}, the change is by a diffeomorphism in the underlying CY,
\begin{equation}\label{eq:shift2}
\delta_I \tilde g_{ij}=\frac{\partial \tilde g_{ij}}{\partial u^I}-\tilde \nabla_i(e^{2A} \eta_{Ij})-\tilde \nabla_j (e^{2A}\eta_{Ii})\,.
\end{equation}
Here $\partial \tilde g_{ij}/\partial u^I$ are the unwarped modes from the
previous section, which are in transverse traceless gauge. The
compensating fields $\eta_{Ii}$ are then fixed by \eq{constr1} and
\eq{constr2}. The physical zero mode $\delta_I \tilde g_{ij}$ is
guaranteed to satisfy $\delta \tilde R_{ij}=0$ separately for K\"ahler
and complex deformations; indeed, it differs from the corresponding
unwarped mode only by a gauge transformation. Notice however that the
zero mode equation is no longer the Lichnerowicz laplacian which is
only valid in harmonic gauge. Rather, one would have to solve the full
\eq{deltaR}. Of course, since we already know $\partial_I \tilde
g_{ij}$, it is simpler to use the constraints to solve for the
compensating fields.

The behavior of the compensators depends on each particular
background, but from the discussion of section
\ref{subsec:comp-effects} we know that they give a nonzero
contribution to the field space metric. In fact, the
correct choice will minimize its
value on a gauge orbit. 
One important consequence of this is that
the metric \eq{warped-G} could mix complex and K\"ahler
moduli. Indeed, a complex structure fluctuation acquires a nonzero
mixed component $\delta_\alpha \tilde g_{a \bar b}$, while the
K\"ahler moduli also have pure components $\delta_r \tilde
g_{ab}$. Therefore, there can be mixed terms of the form
\begin{equation}
G_{\alpha r} \sim \frac{1}{V_W}\,\int d^6y \sqrt{\tilde g_6}\,e^{-4A}\,\big(\delta_\alpha \tilde g_{ab}\,\delta_r \tilde g^{ab}+\delta_\alpha \tilde g_{a \bar b}\,\delta_r \tilde g^{a \bar b}\big)\,.
\end{equation}
This can affect KKLT type~\cite{kklt} scenarios including warping, so
it would be important to understand better the susy structure of the
field space metric.

\section{Analysis of the warped deformed conifold}\label{sec:ks}

In this last section, the previous formalism is applied to construct
the metric of the complex modulus $S$ of the warped deformed
conifold. The warp factor is produced by turning on $N$ units of $F_3$
flux through the A-cycle, and $\beta^{NS}$ units of $H_3$ flux through
the B-cycle.

Let us first note that this problem has a good supergravity limit, in
which $\alpha'$ corrections vanish.  One might worry about this point
because the unit of flux quantization involves $\alpha'$.  However,
one can hold the magnitude of $F_3$ and $H_3$ fixed by scaling up the
number of flux units as one takes $\alpha'\rightarrow 0$.  The only
remaining dependence on $\alpha'$ is in the ten-dimensional Planck
constant, which drops out for $g_s\rightarrow 0$. This is the relevant large $N$
limit in gauge/gravity dualities or compactifications with large hierarchies.

For concreteness, consider a coordinate system where the conifold is
centered around $r=0$. At a distance $r \approx \Lambda_0$ the throat
is glued to a compact Calabi-Yau along the lines described
in~\cite{gkp}. Three regions may then be distinguished:
\begin{itemize}
\item[-] $r \ge \Lambda_0$ corresponds to the transition region into the bulk;
\item[-] $(g_s N \alpha')^{1/2} \le r \le \Lambda_0$ describes a deformed conifold with approximately constant warp factor $e^{-4A} \approx c$;
\item[-] $r \ll (g_s N \alpha')^{1/2}$ is the strongly warped limit of
the deformed conifold, described by the Klebanov-Strassler
solution~\cite{ks}.
\end{itemize}
Notice that in the large N limit $S \ll \Lambda_0^3$.

In the region $r \ge (g_s N \alpha')^{1/2}$ the warp factor variations
may be neglected and the compactification space is a Calabi-Yau
manifold. For small $S$, the bulk contributions are subleading and the
metric $G_{S \bar S}$ is~\cite{conifold}
\begin{equation}
 G_{S \bar S} = \frac{k}{V_{CY}}\,{\rm log}\,\frac{\Lambda_0^3}{|S|}\,.
\end{equation}
Geometrically, the logarithmic dependence follows from a monodromy
argument, and from the dual field theory point of view it is related
to the running of the gauge coupling~\cite{intriligator}. In our
present approach, the compensating fields impose the harmonic gauge
for metric fluctuations, and the computation of the field space metric
is done along the lines of section \ref{subsec:cy}.

On the other hand, a very different behavior may be observed in the
strongly warped region. In~\cite{dg} it was conjectured that the field
space metric including warp effects is
\begin{equation}\label{eq:dg-metric}
 G_{S \bar S}=- \frac{\int e^{-4A}\,\chi_S \wedge \bar \chi_S}{\int e^{-4A}\,\Omega \wedge \overline \Omega}\,.
\end{equation}
Based on this,~\cite{dst} found a new power-like divergence in the metric,
$$
G_{S \bar S}=\frac{1}{V_W}\,\Big(c\, {\rm log}\,\frac{\Lambda_0^3}{|S|}+c'\,\frac{(g_s N \alpha')^2}{|S|^{4/3}} \Big)\,.
$$ 
However, the conjectured form \eq{dg-metric} is not orthogonal to
gauge transformations since $\chi_S$ is harmonic with respect to the
unwarped metric, while the physical fluctuations should be harmonic
with respect to the full 10d metric.

Our aim is to find the correct metric $G_{S \bar S}$ for the strongly
warped conifold using the results of section \ref{sec:kinetic} and
\ref{sec:string}. Before this, we briefly review the KS
solution~\cite{ks}.

\subsection{The  Klebanov-Strassler background}

This is the strongly warped limit of the deformed conifold,
\begin{equation}\label{eq:conif}
\sum_a\, (z^a)^2=S\,.
\end{equation}
The full 10d metric reads~\cite{ks}
\begin{eqnarray}\label{eq:KSmetric}
ds_{10}^2&=&\frac{|S|^{2/3}}{2^{1/3}(g_s N \alpha')}\,I(\tau)^{-1/2}\,\eta_{\mu\nu} dx^\mu dx^\nu+\frac{1}{2^{2/3}}(g_s N \alpha')\,I(\tau)^{1/2}K(\tau) \Big[ \frac{1}{3 K(\tau)^3} \big(d\tau^2+(g^5)^2\big)+\nonumber\\
&+&{\rm cosh}^2\left(\frac{\tau}{2} \right)\big((g^3)^2+(g^4)^2 \big)
+{\rm sinh}^2\left(\frac{\tau}{2} \right)\big((g^1)^2+(g^2)^2 \big)\Big]
\end{eqnarray}
and the warp factor is given by
\begin{equation}
e^{-4A(\tau)}=2^{2/3}\,\frac{(g_s N \alpha')^2}{|S|^{4/3}}\,I(\tau)\,.
\end{equation}
The model is regularized in terms of the UV cutoff $\tau_\Lambda$ defined by $e^{-4A(\tau_\Lambda)} \approx 1$.

A very interesting feature of this solution is that the warped 6d
metric becomes independent of the complex modulus $S$, which only
enters in the redshift factor of the observable energy. This is due to
the fact that in the noncompact limit the $S$-dependence from the warp
factor cancels that of the unwarped metric. As a result, the energy
scales of fluctuations localized in the throat are essentially
controlled by the minimum redshift
$$
e^{A_{min}}\sim\,|S_{min}|^{1/3}=\Lambda\,.
$$ 
In the dual gauge theory this is the statement that there is a mass
gap given by the dynamical scale $\Lambda$.

From this viewpoint, it is not easy to interpret geometric quantities
such as \mbox{$\int \,e^{-4A}\,\chi_S \wedge \bar \chi_S$}, given that
the warped internal metric does not vary under a complex
deformation. Therefore, let us explain how the metric for the S-field
arises.  In this case $\partial_S g_{ij}=0$, so it
is better to work directly with the original expression \eq{hamilton-G},
\begin{equation}\label{eq:Smetric1}
G_{S \bar S}=\frac{1}{4V_W}\,\int d^6y \sqrt{g_6}\,e^{2A}\,g^{ik} g^{jl}\delta_{\bar S} g_{ij}\,\delta_S \pi_{kl}\,.
\end{equation}
Since $\partial_S g_{ij}=0$, we have (suppressing the subindex `S' in $\eta_{Si}$)
$$
\delta_S g_{ij}=-\nabla_i \eta_j-\nabla_j \eta_i\,.
$$ 
Hence the internal metric fluctuation is produced solely by the
compensating field! This contribution is nonzero because a
time-dependent fluctuation in $S$ does modify the 4d piece of the
metric, and this requires non-vanishing compensators. Thus the KS solution
is very good for illustrating the effects of
compensators, since $G_{S \bar S}$ would vanish if they were not taken
into account.

Plugging this metric fluctuation into \eq{Smetric1}, the integrand
becomes a total derivative. Integrating over $\tau$ gives
\begin{equation}\label{eq:Smetric2}
G_{S \bar S}=-\frac{1}{2V_W}\,\big(\int \prod_i\,g^i\big)\, \sqrt{g_6}\,e^{2A}\,\eta_i \,\delta_S \pi^{i \tau}\Big|_{\tau=0}^{\tau=\tau_\Lambda}\,.
\end{equation}
In the remaining of the section we will compute this quantity. Now we
turn to finding the compensating fields, from which the fluctuation
$\delta_S \pi_{ij}$ will be obtained (see \eq{delta-bg}).

\subsection{Compensating fields}

Solving the compensator equations explicitly is a very involved
task. Indeed, \eq{Dconstr} (or, equivalently, \eq{deltaA} and
\eq{constr6d}) gives a system of six coupled second order PDEs, with
coefficients that contain various combinations of (hyperbolic)
trigonometric functions, plus $I(\tau)$ which only has an integral
expression.  Now, the problem is simplified by the fact that in order
to evaluate \eq{Smetric2} only the solutions close to the boundaries
are needed. The approach is then to expand the KS solution near each
boundary, and find the solutions separately in each region after
making simplifying ansatze for the compensators taking into account
the isometries of the background.

Still the problem turns out to be too complicated to allow for an
intuitive understanding of the underlying physics. Instead, we will
consider the so-called \emph{hard-wall} approximation, where the
regular background is replaced by an AdS space with a cut-off at
$r=|S|^{1/3}$ plus boundary conditions to match the known KS
values. The warp factor is taken to be
\begin{equation}
e^{-4A(r)}=\frac{a_0(g_s N \alpha')^2}{r^4}
\end{equation}
where $a_0=2^{2/3}\,I(0)$ is chosen so that at $r=|S|^{1/3}$ this
agrees with the KS warp factor at $\tau=0$. Similarly, the 10d metric
will be approximated by
\begin{equation}
ds_{10}^2=e^{2A(r)}\,\eta_{\mu\nu}\,dx^\mu dx^\nu+e^{-2A(r)}\big(dr^2+r^2\,ds_{T^{1,1}}^2 \big)\,.
\end{equation}
In the hard-wall approximation there is one IR boundary at
$r=|S|^{1/3}$ and the space has a UV cutoff at $r=\Lambda_0$. However,
due to the fall-off of the metric fluctuations at large $r$, only the
IR boundary turns out to contribute to the field space
metric. Therefore we only need to solve for the compensators around
the tip of the conifold.

Before proceeding, let us pause and ask about the validity of this
approximation. The work of~\cite{berg} performed a detailed numerical
analysis of the mass spectrum in the full KS solution without any
approximation in the background. Their results were compared to the
ones obtained in the hard-wall approximation and it is found that,
although the precise numerical coefficients don't agree, both spectra
have the same dependence on the parameters of the problem. Since the
masses depend directly on the kinetic term metric, the hard-wall
method gives the correct dependence on $g_s N \alpha'$ and $S$, while
more work would be required to get the numerical coefficients right.

\vskip 1.5 mm

From \eq{deltaA} and
\eq{constr6d}, the constraint equations that
need to be solved are
\begin{eqnarray}\label{eq:compeq-KS}
&&g^{ij}\,\nabla_i\,\eta_j+2\, g^{ij}\,\partial_i A \, \eta_j=2\,\partial_S A\nonumber\\
&&g^{ij}\,\nabla_i (\delta_S \pi_{jk})+2 \,g^{ij}\,\partial_i A\,\delta_S \pi_{jk}=0
\end{eqnarray}
with
$$
\delta_S \pi_{ij}=-\nabla_i \eta_j-\nabla_j \eta_i-g_{ij}\,(g^{kl}\,\nabla_k \eta_l)\,.
$$ The covariant derivatives here are with respect to the warped 6d
metric $g_{ij}$.

Due to the $SU(2)\times SU(2)$ symmetry, the angular components of the
compensators may be rotated to point in the $\psi$ direction. A radial
compensator is of course needed due to the source term produced by
$\partial_r A$. Then from \eq{compeq-KS} we learn that $\eta_r$ and
$\eta_\psi$ only depend on the radial direction. Notice that at least
two nonzero components are needed to be able to construct a metric
fluctuation orthogonal to gauge transformations. Summarizing, our
ansatz for the compensating field is
\begin{equation}\label{eq:twocom-ansatz}
\eta_i(y)=\big(\eta_r (r),\,\eta_\psi(r),0,0,0,0\big)
\end{equation}
where the last 4 components refer to the coordinates $(\theta_i, \phi_i)$. 

This is admittedly not the most general ansatz; one could find others
with less symmetry.  However, since the kinetic term coefficient
\eq{Smetric1} is the integral of a positive definite quantity, it
seems very implausible to us that a solution with less symmetry could
lead to a smaller result.

Granting \eq{twocom-ansatz}, the system \eq{compeq-KS} then becomes
one second order equation for $\eta_\psi$ and two equations (one first
order and one second order) for $\eta_r$. Concentrating on $\eta_r$
first, the general solution to the first order equation is
$$
\eta_r(r)=\sqrt{a_0}\,\frac{(g_s N \alpha')}{|S|}\,\frac{1}{r}+\frac{c_1}{r^3}
$$ 
Plugging this into the second order constraint sets $c_1=0$. The
role of this compensator is to cancel the contribution of the
nontrivial warp factor; it may be checked that $\eta_r$ is covariantly constant, $\nabla_r \eta_r=0$. This then implies that
$$
g^{kl} \nabla_k \eta_l=0\;,\;\delta_S \pi_{rr}=0\,.
$$

Due to these properties, $\eta_r$ drops out from the second order equation for $\eta_\psi$, and the solution around $r \approx |S|^{1/3}$ is
$$
\eta_\psi(r)=\frac{b_1}{r}\,.
$$ 
The constant $b_1$ is fixed by matching $||\delta_S \pi_{\psi
r}||^2$ at $r=|S|^{1/3}$ to $||\chi_S||^2$ at $\tau=0$, ensuring that
the metric fluctuations are normalized in the same way. This boundary
condition is required because the IR cutoff $r=|S|^{1/3}$ is imposed
by hand. The result is
$$
\eta_\psi(r) \approx k\,\frac{(g_s N \alpha')}{|S|^{2/3}}\,\frac{1}{r}\,,
$$
where from now on we will absorb the dimensionless order one constants
into $k$. The dependence on $(g_s N \alpha')$ and $|S|^{2/3}$ can also
be understood as follows. Since $\delta g_{\psi r}=e^{-2A}\,\delta
\tilde g_{\psi r}$ and $\delta \tilde g$ is independent of fluxes, the
warped metric fluctuation has to be proportional to $(g_s N
\alpha')$. Then $|S|^{-2/3}$ follows from dimensional analysis.

Putting these results together, the compensating field in the
hard-wall approximation is
\begin{equation}\label{eq:comp-final}
\eta_i(y)=\big(\sqrt{a_0}\,\frac{(g_s N \alpha')}{|S|}\,\frac{1}{r},\,k\,\frac{(g_s N \alpha')}{|S|^{2/3}}\,\frac{1}{r},0,0,0,0 \big)\,.
\end{equation}
With these components, the only nonvanishing metric fluctuation is
\begin{equation}
\delta_S \pi_{\psi r}=-k\,\frac{(g_s N \alpha')}{|S|^{2/3}}\,\frac{1}{r^2}\,.
\end{equation}

Naively, one might find it peculiar that the metric variation is an
off-diagonal component, not present in the original Klebanov-Strassler
metric \eq{KSmetric}.  But, as we commented, the 6d part of the
Klebanov-Strassler metric is actually independent of $S$, and the
variation is pure gauge.  Nevertheless it must be non-zero to satisfy
the orthogonality condition.

\subsection{Metric including compensator effects}

To compute the field space metric we need to replace \eq{comp-final}
into the expression \eq{Smetric2},
$$
G_{S \bar S}=-\frac{{\rm vol}(T^{1,1})}{2V_W}\,
  k^2\,r^5\,e^{-4A}\,g^{\psi \psi}\,g^{rr}\,\eta_\psi \,\delta_S \pi_{\psi r}\,,
$$
and then evaluate this at $r=|S|^{1/3}$. The result is
\begin{equation}\label{eq:Smetric-final}
G_{S \bar S}=k\,\frac{{\rm vol}(T^{1,1})}{V_W}\, \frac{(g_s N \alpha')^2}{|S|^{4/3}}\,,
\end{equation}
where we have combined all the order one numerical constants into $k$.
This metric agrees qualitatively with the one found by~\cite{dst}. 

We have arrived to the same functional dependence on $S$ but through a
very different path, by requiring orthogonality with respect to gauge
transformations in the presence of warp and conformal factors. It is
thus instructive to connect our results to the expression
\eq{dg-metric} in terms of the $(2,1)$ form $\chi_S$.

First, the effect of the $\eta_r$ compensator is simply to set
$$
\delta_S A = 0 \;,\;\delta_S g=0\,.
$$ 
In terms of the physical fluctuations, the warp factor becomes
independent of $S$ and the metric fluctuation is traceless. In fact,
both are equivalent by the constraint \eq{constr1}. Then the other
constraint (\eq{constr2}) may be rewritten as
\begin{equation}\label{eq:wharmonic}
 \tilde \nabla^i \big(e^{-4A}\,\delta_S \tilde g_{ij}\big)=0
\end{equation}
which is a warped generalization of the harmonic gauge. The associated 3-form
\begin{equation}
 \chi_S=\,\tilde g^{ln}\,\Omega_{ijl}\,\delta_S \tilde g_{nk}\,dy^i dy^j dy^k
\end{equation}
then satisfies
\begin{equation}\label{eq:dchi}
d \star_6 (e^{-4A}\,\chi_S)=0\,.
\end{equation}
In other words, the effect of $\eta_\psi$ is to shift the original
harmonic $(2,1)$ form by an exact piece so that the ``physical''
$\chi_S$ satisfies \eq{dchi}.

With this constraint, the field space metric reads
\begin{equation}\label{eq:final-metric}
 G_{S \bar S}=- \frac{\int e^{-4A}\,\chi_S \wedge
 \star_6\,\bar \chi_S}{\int e^{-4A}\,\Omega \wedge \overline \Omega}\,.
\end{equation}
The Hodge star is needed because $\chi_S$ is no longer harmonic. 

After having established this, it becomes clearer why we find the same
$1/|S|^{4/3}$ behavior as in~\cite{dst}. The reason is that the
original harmonic form is shifted by an exact piece in order to
satisfy \eq{dchi}, but in the KS coordinates this equation is
independent of $S$. Hence neither the $(2,1)$ form nor the exact
3-form add extra $S$ dependence to \eq{final-metric}.  In fact all of
the $S$ dependence comes from the warp factor, which is proportional
to $|S|^{-4/3}$.  This can be extracted, and the remaining integral
leads to an order one coefficient.  As the integrand is different, its
numerical value is probably different than that of \cite{dst}.  But
since the correct field space metric minimizes a positive definite
inner product, the result must be equal or smaller than
that found in \cite{dst}.

The upshot is that the expression \eq{dg-metric} was qualitatively
correct in this case, however it is not yet clear in what generality
this is true as the argument we just gave depends on special
properties of the KS solution.

\vskip 4mm

To conclude, we would like to point out that, while our approach does
not use supersymmetry, it would be important to understand which are
the implications of these results for the 4d K\"ahler potential. For
instance, while we have proved that \eq{dchi} holds for the conifold,
this may also be valid in compactifications which admit a covariantly
constant spinor in six dimensions. Another possible application is to
computing kinetic terms from compactifications which are not
conformally equivalent to Calabi-Yau manifolds. Such backgrounds may
describe gravity duals of metastable vacua in strongly coupled gauge
theories; see~\cite{kachru} for a recent example. We plan to come back
to this in the future.


\section*{Acknowledgements}

We would like to thank 
D.~E.~Diaconescu, 
A.~Dymarsky, 
J.~Gray, 
S.~Kachru, 
S.~Klevtsov,
S.~Lukic, 
A.~Nacif, 
G.~Shiu,
E.~Silverstein,
and B.~Underwood
for useful discussions and comments. G.~T. would like to thank the
Stanford Physics Department for their hospitality while part of the
project was done.

This research was supported by DOE grant DE-FG02-96ER40959.

\end{document}